\documentclass[aac]{iosart2x}

\newcommand{\systemName}{Socratrees}

\pubyear{0000}
\volume{0}
\firstpage{1}
\lastpage{1}

\begin{document}

\begin{frontmatter}

\title{Socratrees: Exploring the Design of Argument Technology for Layman Users}
\runningtitle{\systemName}

\author{\inits{N.}\fnms{Steven} \snm{Jeuris}\ead[label=e1]{sjeu@dtu.dk}}
\address{Department of Health Technology, \orgname{Technical University of Denmark},
\cny{Denmark}\printead[presep={\\}]{e1}}
\runningauthor{S. Jeuris}


\begin{abstract}

Terms like `misinformation', `fake news', and `echo chambers' permeate current discussions on the state of the Internet.
We believe a lack of technological support to evaluate, contest, and reason about information online---as opposed to merely \emph{disseminating} it---lies at the root of these problems.
Several \emph{argument technologies} support such functionality, but have seen limited use outside of niche communities.
Most research systems overemphasize \emph{argument analysis} and structure, standing in stark contrast with the informal dialectical nature of everyday argumentation. Conversely, non-academic systems overlook important implications for design which can be derived from theory.
In this paper, we present the design of a system aiming to strike a balance between structured argumentation and ease of use. \emph{\systemName{}} is a website for collaborative argumentative discussion targeting layman users, but includes sophisticated community guidelines and novel features inspired by \emph{informal logic}.
During an exploratory study, we evaluate the usefulness of our imposed structure on argumentation and investigate how users perceive it. Contributing to arguments remains a complex task, but most users learned to do so effectively with minimal guidance and all recognized that the structure of \systemName{} may improve online discussion and results in a clearer overview of arguments.
\end{abstract}


\end{frontmatter}


\section{Introduction}

There currently is an unprecedented amount of information available online. This can be used by decision-makers, the civically-engaged, journalists, and researchers alike,
to inform themselves. However, doing so requires navigating and parsing a complicated web of disparate resources (such as news reports,
scientific articles, and social media), which can make it hard to `see the forest for the trees'. Although we have made great advances in how information can be disseminated, there is a lack of technological support to integrate, evaluate, contest, and reason about it~\cite{rahwan2017towards}.

Argumentation online is certainly possible---and commonplace, but typically only adds to the ever-growing torrent of information, leaving behind lengthy disorganized threads, interspersed with random Internet banter.
Since anyone can contribute, verifying the validity of statements found online is extremely time-consuming---time most people do not have or are simply unwilling to give up leisure time for~\cite{flintham2018fakenews}.
Instead, most people decide on a select few news sources (e.g., based on political predispositions~\cite{stroud2008media} or attitudes towards specific issues of interest~\cite{feldman2018explaining}) in which to place their trust and based on which their opinions are formed.
As a result, many information sources act like `echo chambers': discussions of opposing views are inconveniently segregated, and ill-founded ideas can propagate freely as they remain largely unopposed~\cite{mercier2011humans,stroud2008media}.
It is no surprise then, that terms like `misinformation', `fake news', and `filter bubbles' permeate current discussions on the state of the Internet~\cite{wired2017internetbroken}. 

Inspired by \emph{argumentation theory}~\cite{eemeren1999developments,walton2009argumentation}, research on the border of logic, philosophy, and computer science has taken up the challenge to create better tools to disseminate, structure, and analyze rational thought, collectively called \emph{argument technologies}. These target a wide variety of application areas, including
conflict resolution, legal argumentation in law, discussing and documenting design rationale, and sensemaking~\cite{janjua2015survey,scheuer2010computer,schneider2013review}. In this paper, we focus on collaborative, publicly available, web-based technologies with explicit support for interlocutors to view and contribute to argumentative discourse.
While several such technologies exist~\cite{schneider2013review}, they remain a niche outside of research; support for argumentation on mainstream social networks is limited to basic functionality such as up or down voting and conversation threading.
Some researchers argue that the primary obstacle to more widespread adoption is a mismatch between the \emph{informal} nature of online discussions and the highly \emph{formal} structured functionality current argument technologies provide, i.e., \emph{usability}~\cite{paglieri2017plea,schneider2013review}.

We introduce the design of a website for collaborative argumentative discussion, named \emph{\systemName{}}, exploring the delicate balance between structure---an integral part of argumentation---and ease of use.
Prior argument technologies typically rely on complex ontologies to represent claims and relationships between them, e.g., `issue', `position', `challenge', `justification', and `agreement'. In contrast, the user interface introduced in this paper reduces argumentation to but three core concepts for the user to understand: (1) \emph{statements}, that can (2) \emph{support} and (3) \emph{oppose} one another.
Our primary goal is to allow users to form their own opinions by collaboratively aggregating all information relevant to a given statement in one location. As opposed to prior work (e.g., \cite{schneider2007argunet,snaith2010mixed}), our focus is less on analyzing a single argument in great detail and more on providing the necessary structure to represent many competing arguments side by side, without forming judgment as to which one is sound.

Over the course of six weeks, we publicly invited users to try out \systemName{} and engaged 18 users in argumentation by responding to the 128 statements they posted on \systemName{}. A concluding survey was filled out by 14 users. Based on this exploratory study, we discovered unforeseen challenges and important insights which may be of interest to other designers of argument technologies.

\section{Argumentation Theory}

Formal deductive logic falls short when trying to evaluate the quality of arguments expressed in ordinary everyday language (such as political discourse)~\cite{eemeren1999developments,walton2009argumentation}. Practitioners and teachers of logic started challenging the traditional ideals of \emph{validity} and \emph{soundness}
and it was soon recognized that ``[formal logic] had in mind one important subset of arguments, but the realm of argumentation was much broader''~\cite{blair1987argumentation}.
Essentially, the logic of argumentation must be distinguished from formal logic which concerns itself solely with inference/implication; instead, argumentation must be seen as dialectical---a \emph{process} with arguments as a \emph{product} of which both sides need to be investigated, for and against, to see how they interact~\cite{blair1987argumentation}.

Today, \emph{argumentation theory} is an umbrella term for studying the \emph{``production, analysis and evaluation of argumentation''} by adopting both descriptive and normative methods, i.e., by evaluating argumentative discourse empirically, as it occurs, as well as reflecting on the necessary criteria for reasonable argumentation~\cite{eemeren1999developments}.
Given our focus on providing technological support for argumentative discourse, we are particularly interested in a normative approach since it can prescribe the necessary functionality to enable more effective argumentation. Therefore, \emph{informal logic}---defined as \emph{``the normative study of argument''}~\cite{blair1987current}---became a logical choice as the main driving force behind
our design.
Walton~\cite{walton2009argumentation} provides a short introduction to informal logic and presents the following minimal definition of an argument:
\begin{quote}
An argument is a set of statements (propositions), made up of three parts, a conclusion, a set of premises, and an inference from the premises to the conclusion. An argument can be supported by other arguments, or it can be attacked by other arguments, and by raising critical questions about it. 
\end{quote}
Adequacy of premises and inferences in informal logic is less strict than in formal logic. Rather than validity and soundness, Blair and Johnson~\cite{blair1987argumentation} argue for: (1) \emph{acceptability} (start with premises the audience is willing to accept), (2) \emph{relevance} (premises ought to be relevant to the conclusion), and (3) \emph{sufficiency} (premises ought to provide sufficient support for the conclusion).
Premises can either work together to form a \emph{linked argument} or contribute to the conclusion independently, whereby they form a \emph{convergent argument}.
\emph{Argumentation schemes} prescribe commonly used forms of linked arguments in which each premise plays a specific role, e.g., an argument `from expert opinion' or `from analogy'.

With these `rules' of informal logic in mind, there are several ways of attacking an argument~\cite{walton2009argumentation}. First, argumentation schemes have associated \emph{critical questions}, e.g., of an argument from expert opinion you can ask, \emph{``Is the premise consistent with what other experts in the field assert?''}
Considering the acceptability criteria, you can either question one of the premises or \emph{refute} an argument by forming a counter-argument which concludes the opposite. It is worth noting that arguments may attack one another. Lastly, you can argue that a premise is not relevant to the given conclusion, or point out logical fallacies in reasoning.

Based on multiple studies in argumentation theory there is a commonly held belief that humans are inherently bad at reasoning and argumentation, highlighting the need to make `critical thinking' and similar topics part of the core curriculum in education~\cite{scheuer2010computer}.
However, more recent work sketches a less bleak picture of the `layman' arguer~\cite{mercier2011humans,paglieri2017plea}.
When people are sufficiently \emph{motivated}, i.e., when argumentation occurs in a dialogical context as opposed to a decontextualized and abstract task frequently employed in studies, they can evaluate arguments quite accurately.
Furthermore, in contrast to studies that show that people are bad at producing arguments (e.g., by succumbing to \emph{confirmation bias}), they form good arguments when challenged and evidence is made available to them.
In other words, people are better at \emph{evaluating} others' arguments than \emph{producing} their own. Therefore, argumentation is most effective in groups with heterogeneous views~\cite{mercier2011humans}.
\section{Argument Technologies}

Part of the roots of argument technologies can be traced back to some of the earliest work in computer science~\cite{shum2003roots}:
\begin{quote}
... [the] `founding fathers' of today's interactive computing such as Bush and Engelbart envisaged argument construction and analysis as a key objective for the intellectual technologies they were conceiving.
\end{quote}
Engelbart's NLS~\cite{engelbart1968research}, introducing many concepts of personal computing, was designed to `augment human intellect' and Bush's Memex, frequently cited as anticipating the Internet, was introduced `as we may think'~\cite{bush1945memex}.
In later research---still predating the World Wide Web, the essence of hypertext was described as \emph{``a computer-based medium for thinking and communication''}~\cite{conklin1987hypertext}, embodied by systems such as TEXTNET~\cite{trigg1986textnet} and SYNVIEW~\cite{lowe1986synview}.
However, given the extensive scope of technologies supporting reasoning that followed~\cite{scheuer2010computer}, we limit ourselves to reviewing those that are publicly available, web-based, collaborative, and include explicit support for argumentation. 
Schneider et al.~\cite{schneider2013review} have conveniently reviewed exactly such systems (37 in total) and furthermore include a broader discussion of argument technologies. We direct the reader to this excellent review for a complete description of related work.

Here, we provide an overview of the most relevant systems related to \systemName{}, while omitting others which might overlap in functionality but target different use cases (e.g., `IBIS-like' decision-support systems~\cite{conklin1988gibis}, knowledge maps, and public opinion polling tools like Opinion Space~\cite{faridani2010opinion} and Polis\footnote{Polis: \url{https://pol.is/}}).
Specifically, we focus on systems that target `information seeking' dialogues in which the goal is to find or share arguments related to a common topic of interest~\cite{walton2009argumentation}.

Common functionality in such systems is to break up arguments into smaller pieces (going by various names, such as statements, premises, claims, and ideas) and specifying connections between them which describe their relationship (e.g., `supports', `attacks', or `is similar to').
Relations which denote inferences construct the argument.
The resulting underlying data structure is a graph of statements which provides an argumentative overview of a specific discussion.

\subsection{Representation and Structure}
\label{sec:argument-structure}
Although arguments essentially form a graph, they can be visualized in a number of different ways; systems have been classified as \emph{linear}, \emph{threaded}, \emph{container}, \emph{graph}, \emph{matrix}, or a combination thereof~\cite{scheuer2010computer}.
Matrix views are extremely uncommon and therefore will not be discussed here.

\emph{Linear and threaded representations} are most in line with those used in traditional social media, e.g., blogs, and sites supporting conversation threading such as Reddit\footnote{Reddit: \url{https://reddit.com} (2005)} respectively.
Argument technologies extend on this. For example, Rich Trellis~\cite{chklovski2005user} allows a mixture of arbitrary free text with the ability to annotate relations, resulting in a linear but formalized overview of an argument.
Rich Trellis was later extended (yet simplified) to Tree Trellis by reducing the possible types of connections to pro and con and by supporting threading.
Videolyzer~\cite{diakopoulos2009videolyzer} supports threaded discussions by adding and responding to time-anchored annotations in online informational videos (e.g., mark part of the transcript as a claim, express agreement, or indicate quality).

\begin{figure}[b]
\centering
\includegraphics[width=0.7\columnwidth]{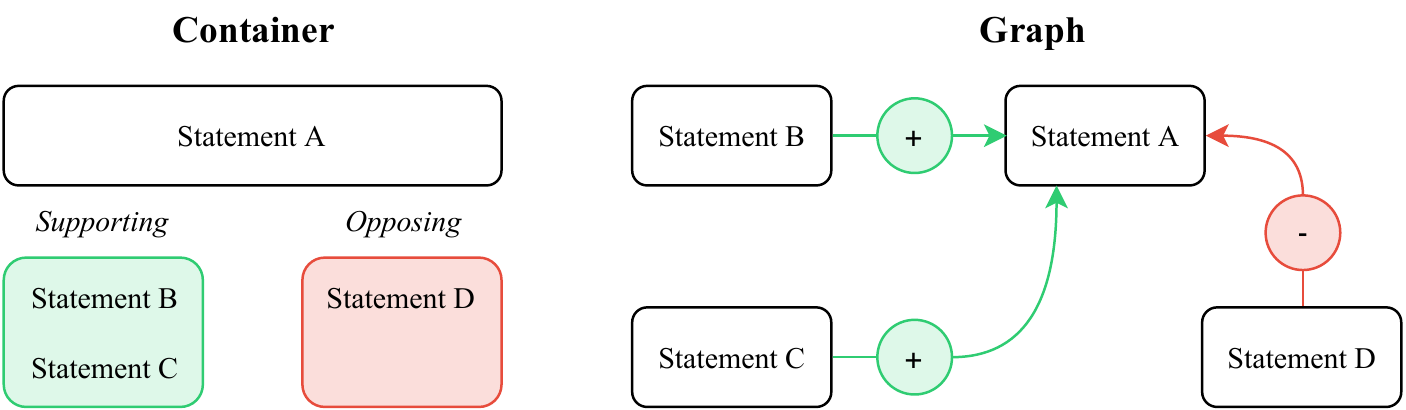}
\caption{Most argument technologies represent arguments as `containers' or `graphs'.}
\label{fig:graph_container}
\end{figure}

\emph{Container- and graph-based representations} are the two most popular approaches in argument technologies. They differ primarily in how connections between statements are visualized. Containers group statements with similar connections to a common `root' statement in a demarcated area, whereas graphs show each connection separately (Fig.~\ref{fig:graph_container}).

An example of a container-based representation is Debatepedia\footnote{Debatepedia: \url{http://www.debatepedia.org} (2007)}, a wiki-based site on which pro and con arguments are grouped together per question on a specific topic.
Most recently, Kialo\footnote{Kialo: \url{https://www.kialo.com} (August, 2017)} (likely the most popular argument technology at the time of writing) was introduced.
Kialo displays all claims in support of/attacking another claim on opposing sides, and frames them in context of an overarching conversation topic.
Although discrete connections are common, some systems provide more granularity. ConsiderIt~\cite{kriplean2012considerit} (now a commercial site\footnote{Consider.it: \url{https://consider.it} (2016)}) aggregates comments on political issues by asking users to register their degree of support on a sliding scale and adding pros/cons to motivate their position. As a result, users can filter pros and cons by degree of support.

\begin{figure}[b]
\includegraphics[width=0.94\columnwidth]{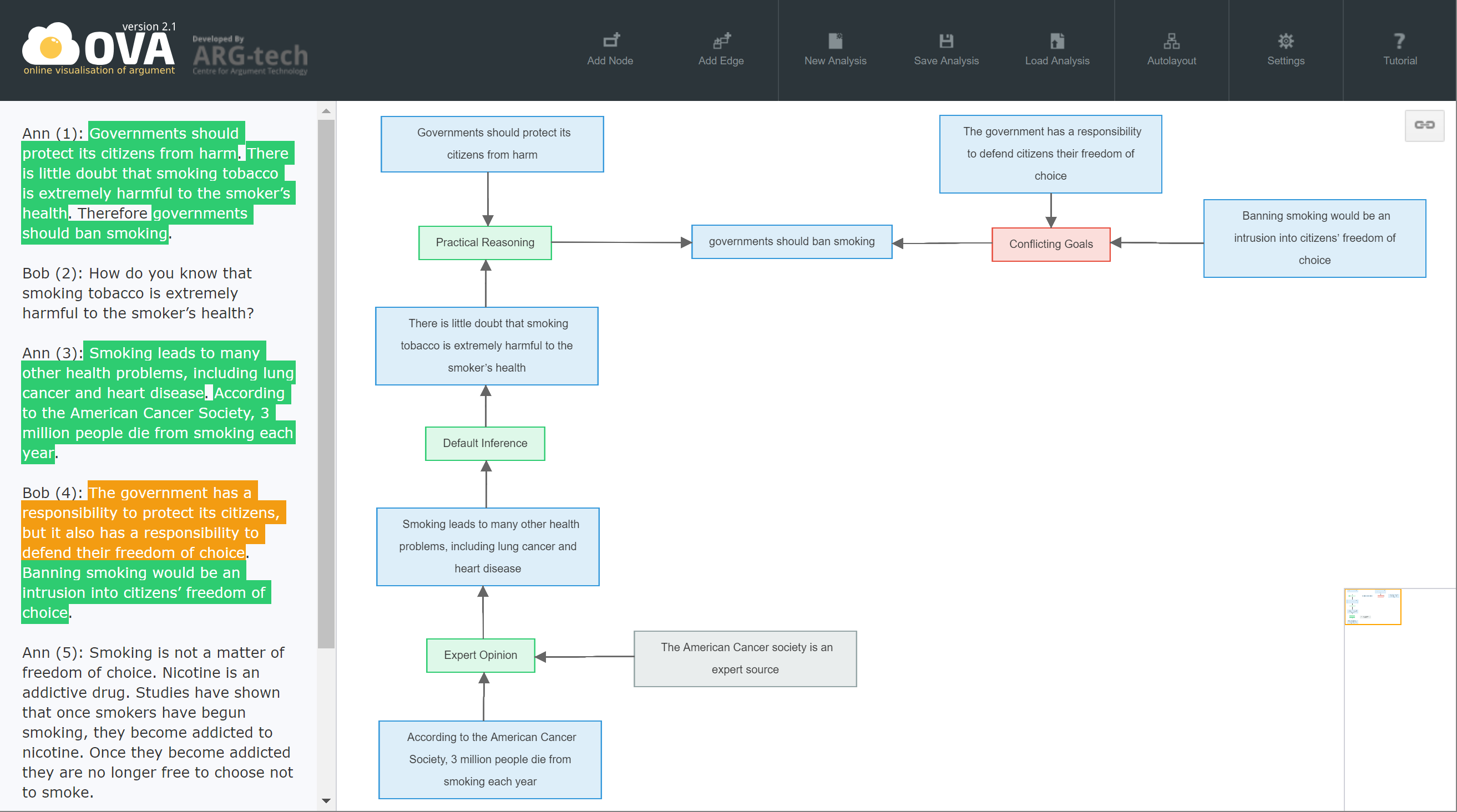}
\centering
\caption{Walton's `smoking dialog' example~\cite{walton2009argumentation}, mapped and visualized in OVA~\cite{snaith2010mixed}. This argument is available on AIFdb: \url{http://www.arg-tech.org/AIFdb/argview/13903}}
\label{fig:ova}
\end{figure}

Container representations provide a clearer, more concise, overview than graphs, at the cost of reducing the types of connections which can be represented.
Therefore, \emph{argument analysis} tools, with their primary focus on formalizing arguments, typically rely on graph representations.
For example, Online Visualization of Argument (OVA)~\cite{snaith2010mixed} loads text (or a web page) side-by-side with a canvas displaying an associated argument map (Fig.~\ref{fig:ova}). Arguments are mapped by highlighting statements and drawing connections in between them on the canvas.
Similar systems are Argunet~\cite{schneider2007argunet}, a desktop tool that supports sharing argument maps online, and the open-source project Arguman\footnote{Arguman: \url{http://arguman.org} (2014)}.
Common functionality is to allow specifying connections in more detail. For example, OVA supports common argumentation schemes such as `practical reasoning' and `expert opinion', relied upon in Fig.~\ref{fig:ova}.
A distinction can thus be made between argument technologies that primarily focus on argument analysis (favoring graph views) and those that focus on information seeking (favoring container views).

In addition to the classification by Scheuer et al.~\cite{scheuer2010computer}, we identify an additional important attribute of argument representation: whether or not a view can be \emph{traversed recursively}, i.e., whether a statement can be made the center of attention by selecting it, thereby hiding all statements that are not directly connected to the selected statement and showing new ones that are. In contrast to displaying a complete argument map (which can typically be panned and zoomed), only a subset of statements are shown at any given time.
Such views are necessary when considering large discussions which may contain cyclical connections and statements that are reused in multiple arguments---a full graph would soon become unintelligible.
Most prior work represents single arguments with limited or no reuse of statements, thereby making recursive views surprisingly uncommon.
However, Debategraph\footnote{Debategraph: \url{https://debategraph.org} (2008)}~\cite{macintosh2009moving} is a graph-based system that supports traversing arguments recursively and 
Kialo supports decomposing arguments into nested hierarchies of pros and cons. Arguments can be explored in great detail by `drilling down' into underlying claims.

\subsection{Social Interaction}
Most argument analysis tools provide limited support for social interaction, other than editing arguments collaboratively.
Conversely, systems with a reduced focus on analysis generally incorporate additional support for collaboration. For example, Cohere~\cite{shum2008cohere}, REASON~\cite{introne2009supporting}, and TruthMapping\footnote{TruthMapping: \url{http://www.truthmapping.com} (2006)} target a more general audience and allow users to express agreement with statements. Similarly, Kialo lets users assess the impact of claims on a 5-point scale.
Such feedback is incorporated to indicate the strength of statements.
Similar to modern social media, argument technologies can trigger notifications when new content is added to arguments a user has contributed to, and content can be moderated (e.g., statements may be reported as spam by users). Kialo seems the most advanced argument technology in regards to supporting social interactions.

\subsection{Infrastructure and Integration}
Several ontologies have been introduced to define the structure of arguments~\cite{janjua2015survey,schneider2013review}, of which the most promising, still in active development, is the Argument Interchange Format (AIF)~\cite{reed2017argument}. AIF enables sharing of arguments across different online services for argumentation, collectively called the `Argument Web'~\cite{bex2013implementing}.
For example, AIFdb\footnote{AIFdb: \url{http://www.aifdb.org} (2012)}~\cite{lawrence2012aifdb} is a public database for arguments, to which the `smoking dialog' represented in Fig.~\ref{fig:ova} was uploaded.
ArguBlogging\footnote{ArguBlogging: \url{http://argublogging.com} (2014)}~\cite{bex2014argublogging} is a browser plugin through which agreement or disagreement can be expressed anywhere online by highlighting text and posting a response to your personal blog and AIFdb.
Similar systems (unrelated to the Argument Web) are rbutr\footnote{rbutr: \url{http://rbutr.com} (2012)} and hypothes.is\footnote{hypothes.is: \url{https://web.hypothes.is} (2013)}, providing a comparable infrastructure to support `open annotation' anywhere on the web.
Ontologies are key to automating argument mining, integration, and evaluation~\cite{janjua2015survey,reed2017argument} (e.g., through the use of artificial intelligence), all of which are considered outside of scope in this paper.

\section{Design Principles/Goals}
Before presenting \systemName{}, we will introduce key design principles which have influenced our design and contrast them with prior work.

\subsection{Transparency First---Inspire Critical Thinking}
\label{principle:transparency}
Technology should augment rather than replace human judgment~\cite{flintham2018fakenews}.
In line with this recommendation, our goal is not to prescribe what is true or false (i.e., to be a fact finding tool), but to provide \emph{transparency} to arguments and to \emph{inspire critical thinking}.
Rather than presenting single arguments in great detail, we aim to provide an easily digestible overview of all relevant information in relation to a specific statement. Competing arguments thus live side-by-side and it is up to users to interpret them and draw their own conclusions.
In other words, our main goal is not supporting `argumentation' per se, but providing a record of the collaborative thought process in order to aid individual human reasoning.
Such overviews facilitate \emph{distributed sensemaking}~\cite{fisher2012distributed}, and \emph{group reasoning} in which argumentation theory predicts truth to win out~\cite{mercier2011humans}.

\subsection{Help Finding Common Ground}
\label{principle:commonground}
Argumentation theory describes different types of dialog with differing goals: e.g., in a \emph{persuasion} dialog the goal is to convince the other party, and in a \emph{deliberation} dialog the goal is to decide on the best available course of action~\cite{walton2009argumentation}.
Argument analysis systems (as discussed in related work) primarily target persuasion dialogs.
In contrast, \systemName{} targets \emph{information-seeking} and \emph{inquiry} dialogs, in which the goal is to exchange information and find and verify evidence respectively.
By sharing statements in relation to one another (as supporting or opposing) and allowing users to express agreement with each, an overview becomes available of how well-supported statements are, the different reasons for believing or not believing in them, and how popular these are. Strong arguments on either side of a discussion indicate `common ground', whereas statements on which opinions are divided reveal true points of contention.

\subsection{Conducive to Large-scale Discussions}
\label{principle:scale}
Information-seeking and inquiry dialogs are cooperative in nature, as opposed to persuasion dialogs which tend to be adversarial, as summarized by the philosopher Neurath~\cite{neurath1940universal}:
\begin{sloppy}
\begin{quote}
Debaters on comprehensive scientific problems are ... like lawyers who have to take a side. Each of them intends to strengthen his own arguments and to weaken the arguments of the aggressor---but no judge is in the chair. ... Finally we find ourselves all together in the same ship and are co-operating even when we think we are fighting one another.
\end{quote}
\end{sloppy}
To better support the cooperative nature of argumentation, we explore interaction techniques for large groups of users to collaboratively contribute to discussions comprising numerous statements, without hindering one another, and while retaining a suitable overview for people arriving later to, or having missed part of, the discussion.

Based on our review of web-based argument technologies, we identify three core challenges to supporting \emph{large-scale} discussions online: (1) dealing with \emph{digressions} in order to ensure \emph{focused} argumentation, (2) supporting users to \emph{explore arguments at their own pace}---without being overwhelmed by expert accounts, and (3) enabling/encouraging \emph{statement reuse} in order to eliminate redundant discussions and to capitalize on prior knowledge.
TruthMapping has a built-in mechanism to deal with digressions, and Kialo supports focused exploration of arguments by breaking them up into concise claims. However, statement reuse in prior systems is rare or simply unsupported. Kialo supports reusing claims through the use of `symlinks', but in practice this feature is hardly used. We observe that for statements to be reused, it is essential that they are \emph{`free of context'}, i.e., that they can be interpreted outside of the context in which they were introduced. Prior systems do not enforce this.

\subsection{Inclusiveness}
\label{principle:inclusiveness}
Prior academic work has mostly enforced argument structure, requiring knowledge of argument analysis, at the cost of \emph{usability}~\cite{paglieri2017plea,schneider2013review}.
We believe that given a more suitable medium, anyone can contribute meaningfully to argumentation online~\cite{mercier2011humans,paglieri2017plea}.
Users should not be expected to know argumentation theory in order to start using argument technologies.
Similar to Cohere~\cite{shum2008cohere} and the non-academic systems TruthMapping and Kialo, we target a more general audience. Our goal is to find a suitable balance between imposing structure and ease of use.
In addition, we aim to be non-discriminatory; it should always be possible for minorities and repressed groups to share unpopular or controversial beliefs.
\section{\systemName{}}

\begin{figure}
\includegraphics[width=\linewidth]{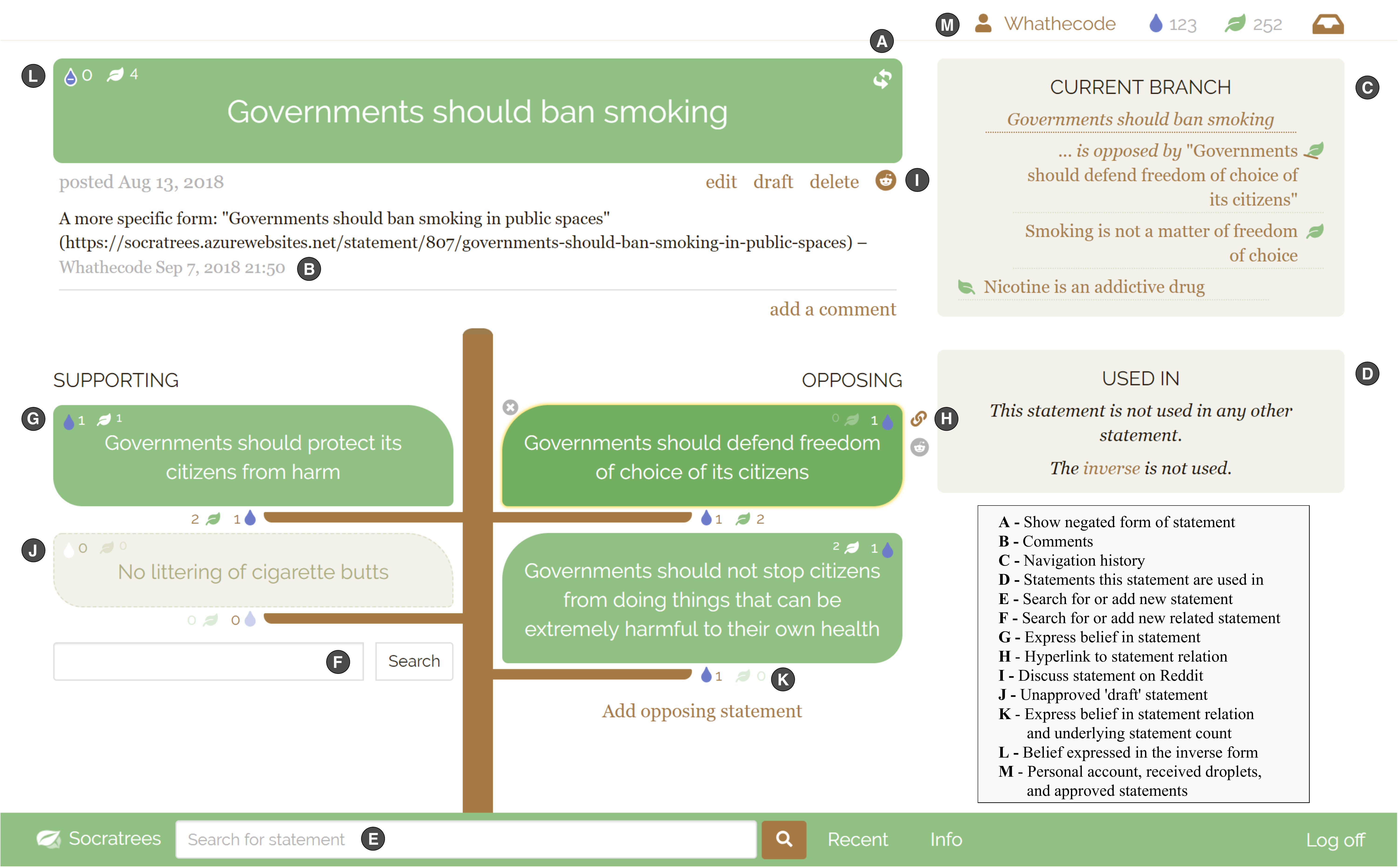}
\caption{The main user interface of \systemName{}, presenting part of the `smoking dialog' also displayed in Fig.~\ref{fig:ova}.}
\label{fig:socratrees-smoking-dialog}
\end{figure}

\systemName{}\footnote{Socratrees: http://socratrees.wiki/} 
(Fig.~\ref{fig:socratrees-smoking-dialog}) is a collaborative, web-based, argument technology, targeting a general, non-expert, audience. The main user interface supports outlining arguments into hierarchies of supporting and opposing statements and relies on a container representation which can be \emph{traversed recursively} to navigate between statements (as described in Section~\ref{sec:argument-structure}).
This supports \emph{focused} argumentation and helps dealing with \emph{digressions}.
Kialo is the only other system combining a container representation with recursive navigation.
In addition, \systemName{} introduces two novel features to structure arguments: (1) statements can be represented both in a normal and negated form, and (2) statement relations (one statement supporting/opposing another) are considered statements themselves which further supporting/opposing statements can be added to.

First, we present the \emph{structure} of \systemName{} and our motivation behind it in more detail, including how \emph{navigation} is supported.
Second, we explain the chosen \emph{visual metaphor} and how we intend to evolve it over time.
Lastly, we present community features and how we believe they can be used to inspire and scale up argumentation.

\subsection{Structure and Navigation}
We have chosen to only support two types of relations between statements---\emph{supporting} and \emph{opposing}---similar to earlier argument technologies targeting a layman audience.
We find this to be a suitable balance between structure and ease of use. Additional semantics are implicit and we trust users to be able to infer more concrete relations based on context.
For example: \emph{``Pineapple does not belong on pizza''} is supported by \emph{``Italians don't put pineapple on pizza''} and opposed by \emph{``Hawaiian pizza is very popular in Australia''}. An explicit analysis of these arguments is unnecessary to understand the two points made.
We have opted to label relations as `supporting/opposing', in contrast to the more common `pro/contra' and `supporting/attacking'. Early feedback indicated this alternative wording inspired animosity and one-sided thinking, counter to our design principles (\emph{`finding common ground'}). For the same reason, supporting and opposing statements are depicted using the same color, in contrast to prior work which mostly relies on green and red respectively. Although differing colors make sense in light of the `redundancy gain' principle in Human-Computer Interaction (HCI), we find it more important to highlight what statements have in common rather than what sets them apart; all relevant statements contribute to the discussion in a meaningful way.

This similarity is further emphasized by the following observation: when considering the negated form (inverse) of a statement, all supporting statements become opposing statements, and vice versa. For example: \emph{``Governments should defend freedom of choice of its citizens''} (as highlighted in Fig.~\ref{fig:socratrees-smoking-dialog}.H) becomes a supporting statement when considering that \emph{``Governments should \emph{not} ban smoking''} instead of \emph{``Governments should ban smoking''}.
\systemName{} exploits this fact by supporting both a normal and negated form for statements. By clicking the `inverse' icon (Fig.~\ref{fig:socratrees-smoking-dialog}.A), a negated form of the statement is shown and the supporting and opposing statements are swapped.
A default \emph{``(not)''} text is prepended to the normal form of statements, but users can specify a custom, more suitable, text for the negated form.
This has two major advantages which reinforce our underlying design principles: (1) better \emph{statement reuse} since the normal and inverted form of statements are structurally identical, and (2) discussions on unpopular or controversial ideas (e.g., holocaust denial) are directly linked to the arguments against.
This makes suppressing them nonsensical as it would also eliminate the arguments in support of what the majority of society believes to be factual or `true'; this supports our `\emph{inclusiveness}' design principle.
To our knowledge, \systemName{} is the first system supporting both a normal and negated form for statements.

\begin{figure}
\includegraphics[width=0.5\linewidth]{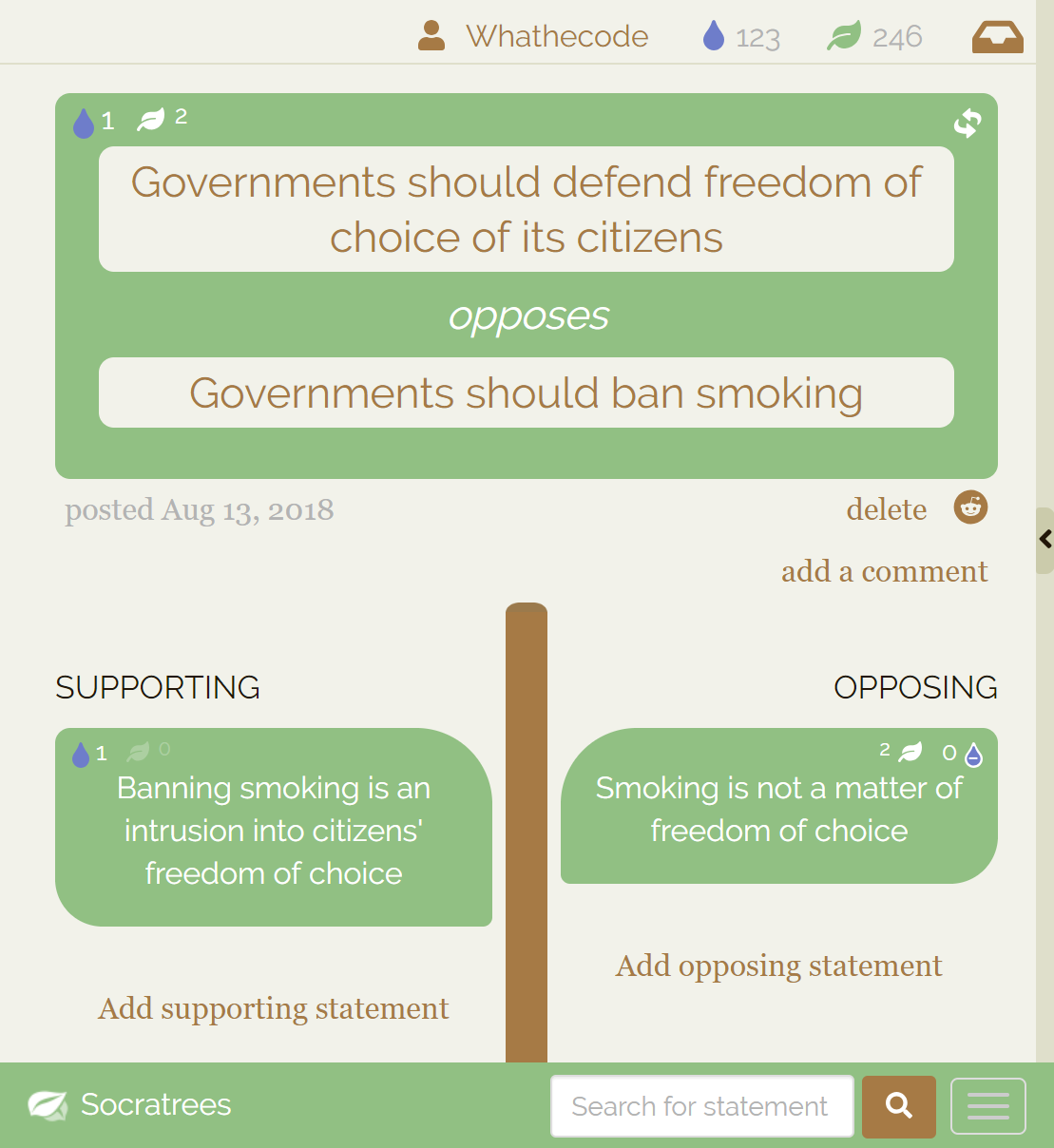}
\caption{The statement relation which is shown when following the branch of the highlighted statement in Fig.~\ref{fig:socratrees-smoking-dialog}.}
\label{fig:relation-statement}
\end{figure}

The last mechanism available in \systemName{} to structure arguments is related to the requirement of \emph{relevance}.
Arguing whether a statement is relevant is an essential, common, part of argumentation, supported in traditional argument analysis tools by applying argumentation schemes and \emph{warrants} to statement relations. Systems targeting a more general audience like Kialo, Consider.it, and TruthMapping do not provide support for this, presumably to sidestep this additional complexity in favor of ease of use.
We try to find middle ground by repurposing the basic building block of argumentation---the statement---to represent statement relations as well.
Concretely, when adding a statement $A$ as a supporting/opposing statement to statement $B$, the system introduces this relation as the statement \emph{``$A$ supports/opposes $B$''} with the inverse form \emph{``$A$ does not support/oppose $B$''}. Similar to how underlying statements can be accessed recursively, the relation statement can be accessed by clicking the stalk which connects the statement to the argument `tree' and further underlying statements can be added to it.
For example, following the relation of the statement highlighted in Fig.~\ref{fig:socratrees-smoking-dialog} leads to the page shown in Fig.~\ref{fig:relation-statement}.
The newly created statement thus acts as the starting point for a discussion on whether or not the given statement is relevant in relation to the other (\emph{``Does A support B?''}), reusing concepts the user is already familiar with.
It can also be used to construct \emph{linked arguments} (statements working together to reach a conclusion).
Further stalks within the statement relation are hidden since reasoning about them becomes overly complicated (\emph{``Does C support A supporting B?''}) and the need to discuss them has not arisen during early testing.

For these structural features to work as intended, we introduce three requirements statements need to comply with:

\noindent \textbf{Be concise}:
Keeping statements concise keeps discussions focused, makes statements more likely to be reused, and makes interpreting their negations easier. When possible, conjuncts should be avoided by splitting statements into two. As argued in the design of Tree Trellis~\cite{chklovski2005user}, this comes at the expense of expressiveness, but also improves the ability to elaborate on arguments without restructuring them, and ensures ancillary points are not mixed with more central ones. To enforce this, statements have a maximum of 120 characters; a limit which did not inhibit us from expressing our own arguments during early testing.

\noindent \textbf{Aim to be free of context}:
It should be possible to interpret a statement outside of the context within which it was introduced. For example: \emph{``Wind turbines are more effective''} makes sense in relation to a statement about solar panels, but takes on a different meaning when used in the context of nuclear energy.
To ensure effective reuse of statements, statements should thus be able to stand on their own and eliminate \emph{indexical} references (e.g., `I', `also', `this', and `that').

\noindent \textbf{Not be phrased as questions}:
Questions do not contribute concrete information to the structure of arguments, and rather indicate specific information is lacking or missing. In essence, the answers to questions are what constitute valid statements on \systemName{}.
This last requirement is controversial due to the unmistaken value of \emph{critical questions} in discussions. However, critical questions can still be posted as \emph{comments} on statements (Fig.~\ref{fig:socratrees-smoking-dialog}.B).

Navigating arguments is supported by traversing statements recursively. At any given time, only a single statement with its supporting and opposing statements is displayed. Opening a different statement is done by clicking on it.
While this ensures \emph{focus} and allows users to explore parts of an argument most relevant to them (in line with our design principles), this becomes disorienting when following many links.
To see which statements you followed to reach your current position in a tree, a `current branch' section is displayed in the sidebar (Fig.~\ref{fig:socratrees-smoking-dialog}.C).
Previously visited statements can be revisited by clicking on them, without erasing the stored navigation path. E.g., Fig.~\ref{fig:socratrees-smoking-dialog} shows the path followed to reach the previously visited statement \emph{``Nicotine is an addictive drug''}. Only when diverging from the presented path, the stored navigation path is overridden.
The sidebar also shows which other statements the current statement is used in (Fig.~\ref{fig:socratrees-smoking-dialog}.D).

New statements can be added after having searched for existing statements first (Fig.~\ref{fig:socratrees-smoking-dialog}.E). An intermediate screen shows search results and an option to proceed with adding the statement in case the desired one is not found. Adding supporting and opposing statements works in a similar way (Fig.~\ref{fig:socratrees-smoking-dialog}.F). Importantly, the search results page also allows users to view and select the normal or negated form of statements, since the form to choose depends on the argument you want to make.
Enforcing `search first' further encourages statement reuse.
Statements are anonymous in order not to discourage people from expressing unpopular beliefs. However, comments are linked to usernames.

\subsection{Visual Metaphor}
After considering other visual metaphors (`branching rivers' and `neural networks'), we eventually chose to represent statements and their supporting and opposing arguments as \emph{leaves on a tree}.
This is a visually rich metaphor that everyone can relate to and evokes a sense of tranquility which might counteract the adverse nature of argumentation.
When you agree with or believe in a statement you can `give it water', represented as droplets (Fig.~\ref{fig:socratrees-smoking-dialog}.G).
Although our current exploitation of this metaphor is limited, we chose to represent statements as leaves since the physical properties of a leaf correlate nicely with those we need to represent.
Statement relevance (the statement relation) is presented by the stalk leading up to the leaf. The stalk might be `broken' when a statement is deemed irrelevant as determined by underlying statements. Leaves in turn can have different colors depending on their `health'. Once we have explored the properties of what makes a good or bad statement in more detail during an extended evaluation we intend to include such visualizations.

\subsection{Community Features}
Although our primary focus in this first iteration of \systemName{} was to explore an alternative structure for argumentative discourse, we implemented basic communication, notification, statistics, and moderation features which provide a strong basis for future work.

We recognize that the potential of \systemName{} lies not in replacing existing discussion platforms, but \emph{augmenting} them.
Therefore---by design---we provide limited support for unstructured argumentation; free form \emph{comments} can be added to each statement (Fig.~\ref{fig:socratrees-smoking-dialog}.B) to add unstructured thoughts, but their primary purpose is to question, clarify, and add related resources. Lengthy discussions are discouraged\footnote{This is inspired by comments on Stack Exchange (\url{https://stackexchange.com/}) which are seen as transient. Valuable comments should be incorporated in the main content of the site (questions and answers).}.
We count on external websites, less restrictive in form and embodying richer communities, to link to content created on \systemName{} and extend on discussions there.
To enable this, each statement has a representative URL (e.g., \emph{``/statement/657/governments-should-ban-smoking''}), and users can link directly to individual comments and statement relations which causes the linked content to be highlighted (Fig.~\ref{fig:socratrees-smoking-dialog}.H). Furthermore, for our exploratory study we integrated with a Reddit discussion board. Clicking a Reddit icon linked to the creation of a new Reddit post with a matching title and a link back to the specific content on \systemName{} (Fig.~\ref{fig:socratrees-smoking-dialog}.I).

The first few statements new users post are posted as \emph{drafts} (Fig.~\ref{fig:socratrees-smoking-dialog}.J). Draft statements are under review and cannot be added to or used as related statements until they are approved.
We implemented this feature so that moderators can provide feedback in comments to new users in case their statements do not comply with site guidelines.
Moderators can also turn problematic statements into drafts.
For now, moderators are predetermined, but we intend to adopt a reputation-based \emph{distributed moderation} model similar to Stack Exchange~\cite{mamykina2011design}.

Users can add `droplets' to leafs and stalks (statement relations) they believe in. The accuracy of statements can thus be judged independently of their relevance. The total number of users agreeing with a statement or its relevance is displayed next to the droplet.
In addition, a leaf icon indicates the total number of underlying statements (Fig.~\ref{fig:socratrees-smoking-dialog}.K).
To express belief in the negated form of a statement the inverse view needs to be loaded first. This forces users to view underlying statements prior to expressing their disbelief---the equivalent of `down voting' in traditional social media. We hope this may inspire users to clarify their disagreement, which we deem more constructive than mere down voting.
The droplet icon is grayed out when you have not formed any opinion and shows a `minus' when you have expressed belief in the inverse form of the currently shown statement (Fig.~\ref{fig:socratrees-smoking-dialog}.L). The result is an overview of all the statements you have formed an opinion on.

Adding droplets also acts as a subscribe mechanism; users receive notifications about changes to any of the statements they have expressed belief or disbelief in.
We feel expressing your opinion should go hand in hand with a willingness to participate in a discussion about it, and have therefore opted to combine these two features into one.
Lastly, a top bar shows an inbox for notifications, in addition to a link to the user's account, the number of times people have expressed agreement with statements added by the user (regardless of form), and the number of approved statements posted by the user (Fig.~\ref{fig:socratrees-smoking-dialog}.M).
\section{Exploratory Study}
Early mock-ups based on the analysis of Reddit threads proved to be too limiting to evaluate the feasibility of our proposed structure for argumentation. To explore how statement negations and our representation of statement relevance influence structuring and interacting with arguments, a functional system had to be built---these features are missing in prior work. In this sense, the presented system can best be described as an \emph{`interactive sketch'}~\cite{greenberg2008usability}:
\begin{quote}
... early designs can also be implemented and maintain their sketch-like properties, ... to explore the ideas behind highly interactive systems. When systems are created as interactive sketches, they serve as a vehicle that helps a designer make vague ideas concrete, reflect on possible problems and uses, discover alternate new ideas and refine current ones.
\end{quote}
Thus, the main purpose for implementing \systemName{} was to \emph{``evaluate the provided functionality to structure arguments''} as specified in the preregistration for this study\footnote{Preregistering studies is recommended by Cockburn et al.~\cite{cockburn2018preregistration}. The preregistration for this study is available on AsPredicted: \url{https://aspredicted.org/blind.php?x=7p26xn}}.
This evaluation is a first step towards answering the overarching research question: \emph{``how to strike the right balance between introducing functionality for structured argumentation and usability in order to obtain more widespread adoption of argument technologies?''}

The author of this paper analyzed three arguments in detail and added them as statements on \systemName{}: a popular blog post on a controversial topic in software development (having received over 100,000 views and 100 comments), a discussion on whether or not \systemName{} improves online discussions, and a smaller argument about smoking taken from literature on informal logic~\cite{walton2009argumentation}, part of which is represented in Fig.~\ref{fig:socratrees-smoking-dialog} and Fig.~\ref{fig:relation-statement}. These and other smaller arguments were added successfully, based on which we conclude \systemName{} is capable of representing a diverse set of real-world arguments.

Next, we evaluated \systemName{} over the course of six weeks involving 32 users. We wanted to assess to which degree users agree our proposed structure has the potential to achieve the design goals outlined earlier: (\ref{principle:transparency}) \emph{transparency first---inspire critical thinking}, (\ref{principle:commonground}) \emph{help finding common ground}, (\ref{principle:scale}) \emph{conducive to large-scale discussions}, and (\ref{principle:inclusiveness}) \emph{inclusiveness}.
The website was announced to friends and colleagues in person, on social media, and several public websites (e.g., subreddits dedicated to argumentation and critical thinking).
238 unique users visited the site, out of which 59 signed up to participate in private beta and were granted access. Given the lack of extensive moderation features, we restricted public access so that only users with access could view statements and contribute to the site. 27 users did not visit the site after having received access, based on which we conclude that 32 users in total participated (to some degree) in private beta. This includes the author of this paper who acted as a moderator (approving draft statements and commenting on them), in addition to acting as an ever-present `active' community member who engaged new users in discussions.
Without moderation, users would be able to add statements that do not comply with site guidelines, which as discussed earlier are essential for our structural features to work as intended.
Lacking a larger community that has a thorough understanding of site guidelines, we believe this was the only reasonable way to mimic an active community and what we eventually hope to enforce by applying a distributed moderation model similar to Stack Exchange~\cite{mamykina2011design}.

\begin{figure}
\includegraphics[width=0.75\columnwidth]{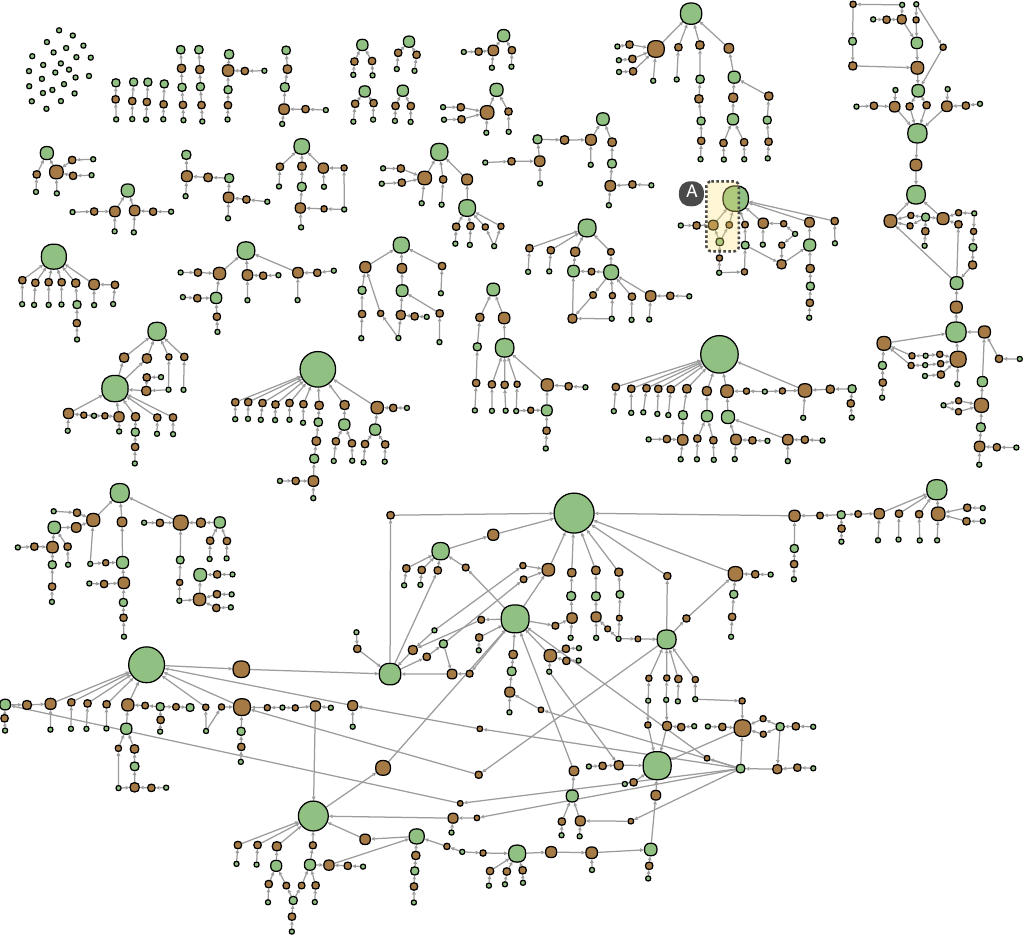}
\caption{Overview of all statements (green) and statement relations (brown) posted on \systemName{}, indicating interlinked arguments.}
\label{fig:argument-topologies}
\end{figure}

19 users added 374 statements and 371 statement relations, so 13 users only browsed the site.
Because the evaluation required participation by the author, the majority of statements were added by him, including the three arguments which were analyzed in detail: 246 statements and 286 relations. Thus, the remaining 18 users added 128 statements and 85 relations.
Fig.~\ref{fig:argument-topologies} shows an overview of all connections between statements, laid out so that individual arguments and connections between them can be recognized.
This gives an impression of argument complexity and why it was necessary to evaluate this structure using an interactive system.
45 statements (12\%) were used more than once (as supporting/opposing).
Negated forms of statements contributed to statement reuse and lead to a particularly interesting use case: at times, both forms of a statement were added to opposing sides of an argument, thereby highlighting key points of disagreement (e.g., \emph{``Global warming is/is not man-made''} supports/opposes \emph{``Climate change is man-made''} in Fig.~\ref{fig:argument-topologies}.A).
The average text of \emph{approved} statements was 58 characters long ($N=367$, $SD=24.0$, $Min=16$, $Max=119$).

All active users received an online survey at the end of the study, which was filled out by 14.
Our survey data shows a wide variety of users: the median age was 28.5 ($SD=11.3$, $Min=23$, $Max=64$), 8 male and 6 female, with backgrounds such as data science, HR consulting, IT, and project management.
On a 5-point Likert scale from \emph{``Not at all familiar''} to \emph{``Extremely familiar''}, nobody specified familiarity with argumentation theory higher than \emph{``Neutral''} ($\mu=2.1$, meaning \emph{``Not familiar''}).
The median number of hours users reported having used the site was 2 ($SD=2.8$, $Min=0.3$, $Max=10$).

\begin{figure}[b]
\includegraphics[width=0.7\columnwidth]{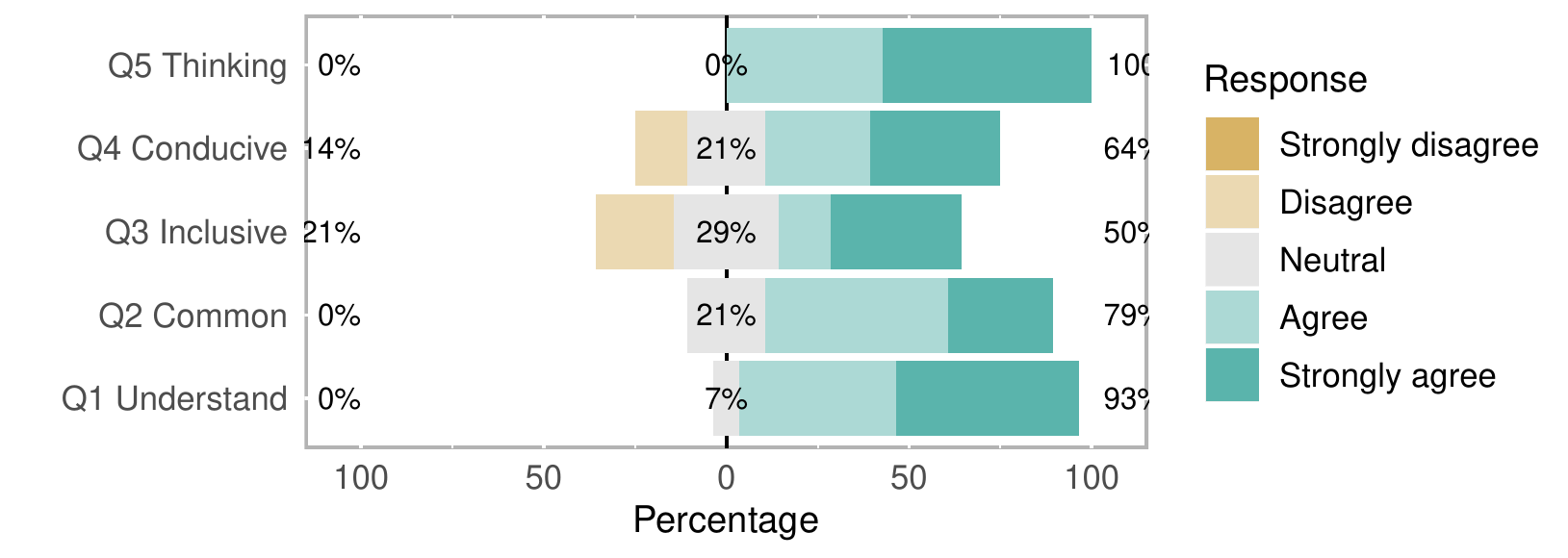}
\caption{Survey results assessing design goals success.}
\label{fig:survey-likert}
\end{figure}

Five central survey items made claims about the potential of \systemName{} and had to be rated on a 5-point Likert scale from \emph{``Strongly agree''} to \emph{``Strongly disagree''}.
Each scale provided the option to enter free form feedback to clarify the provided rating.
Users were instructed as follows:
\emph{``We are interested in your assessment of how well the design of \systemName{} might support argumentation online. To this end, we are primarily interested in feedback related to the core structure represented in the user interface.''} To make sure users understood what was implied by `core structure', a short summary of structural features was included. Each question started with \emph{``\systemName{} ...''}
\begin{description}
    \itemsep0em 
    \item{\textbf{Q1}--design goal 1} \emph{``... can help understanding arguments''} 
    \item{\textbf{Q2}--design goal 2} \emph{``... can help finding common ground''} 
    \item{\textbf{Q3}--design goal 4} \emph{``... is non-discriminatory, i.e., can represent the opinion of minorities and repressed groups''} 
    \item{\textbf{Q4}--design goal 3} \emph{``... can host conducive, noninflammatory, discussions''} 
    \item{\textbf{Q5}--design goal 1} \emph{``... inspires critical thinking''} 
\end{description}
The distribution of responses (Fig.~\ref{fig:survey-likert}) indicates most users believe \systemName{} has the potential to achieve the goals we originally laid out in our design principles. Free form feedback was aggregated using thematic analysis and we will present a representative quote for each individual finding next.

(\textbf{Q1}) 11 users elaborated that the division of supporting/opposing statements provides a \emph{``convenient''} and \emph{``quick overview''} which \emph{``makes it easier to understand arguments''} and \emph{``figure out why two opinions differ.''}
However, one user noted that \emph{``the writer's definition of a term [might be] inconsistent with the reader's''}, potentially leading to conflicts in understanding.

(\textbf{Q2}) Most users indicated that \emph{``[b]y separating arguments into small and separate discussions, it becomes easier to see where you agree/disagree''} and you \emph{``are more likely to read both sides''}. They noted that you are more likely to agree on a single statement than on a whole argument; \emph{``people can agree on having different value-sets.''}

(\textbf{Q3}) A common concern was \systemName{} requires a \emph{``special skill set only common in people with higher educations''}, e.g., having a \emph{``sense of how important details can be in a [statement]''}.
And, while it \emph{``allows for expressing [anonymous] opinions, regardless of their popularity''}, two users expressed concerns about \emph{``counterfactual''} statements. Allowing widespread, uncontrolled, discriminating statements \emph{``would run counter to the goal of being non-discriminatory.''}

(\textbf{Q4}) Four users shared the belief that \emph{``[t]he structure of the platform leaves little room for personal attacks and puts the topic at hand at the core of any discussion.''} Three others argued that discussions are inherently \emph{``emotional''} and will always elicit \emph{``strong reactions''}. \systemName{} can \emph{``temper [this] but can't eliminate it''}.

(\textbf{Q5})
Users overwhelmingly agreed that \emph{``[s]tructured argumentation encourages ... users to understand the relationship between statements.''} \systemName{} \emph{``forces''} users to \emph{``think before [they] write''}, make the implicit explicit, and \emph{``to be very precise in the relation between [statements]''}. However, you \emph{``have to spend a tremendous amount of energy to molding your statements to this structure and all its rules''} and some people \emph{``will never have the interest in exploring views that are different to their own''}.

Lastly, we asked users what would encourage them to contribute more to a site like \systemName{}. Their primary request was a \emph{``larger community''} with \emph{``topics that are relevant to current issues.''} Users want to see that \emph{``[their] statements are making a difference''} and to this end envisioned features such as sharing statements from external websites (supporting \emph{``evidence-based links''}), enhanced reward mechanisms (gamification), and identifying like-minded individuals based on tags for \emph{``ideologies and beliefs''}.
The lack of such features left some users wanting for a \emph{``purpose''} and \emph{``clear goal''}.
\section{Discussion---Challenges}

The results of our study indicate that the structure for argumentation introduced in \systemName{} shows much promise. Even though argumentation remains a complex task, compared to how it currently takes place on social media, users learned to appreciate and understood the trade-offs made, as indicated by our survey results. When asked what would encourage users to contribute more, they did not request a simplified design. Rather, they envisioned scenarios in which a broader community would adopt the given technology.
However, the design was not without flaws. The author of this paper used \systemName{} extensively to structure his own arguments and moderated all content added to the site, based on which a thorough understanding of limitations to the current design arose. In addition, multiple users provided feedback via private communication channels and on the Reddit discussion board, which also contributed to a thorough understanding of open challenges.

\paragraph{Transitioning from Unstructured to Structured Content}
Comments were underused and not fully understood by users. Even though they were designed as a staging ground for thoughts that could not yet be phrased as concrete statements, some users---unaware of this feature---initially expressed a desire to add less-structured argumentation. Comments thus need more exposure and additional mechanisms need to be put in place to go from unstructured to fully-structured argumentation, e.g., incremental and system-assisted formalization mechanisms~\cite{shipman1999formality}.

\paragraph{Reasoning about Relevance is Hard}
Adding statements to statement relations in order to discuss relevance is confusing and particularly error-prone (i.e., adding statements to the statement relation depicted in Fig.~\ref{fig:relation-statement}). While we maintain this is an essential feature, it requires keeping track of three statements and two statement relations in parallel, which can be cognitively demanding.
Furthermore, it is not always clear whether two statements should be added as one arguing for the relevance of the other, vice versa, or both.
For example, \emph{``All questions seek knowledge''} supports \emph{``There are no stupid questions''}, which is relevant because \emph{``Seeking knowledge cannot be stupid''}. But, an alternate phrasing would be \emph{``Seeking knowledge cannot be stupid''}, which is relevant because \emph{``All questions seek knowledge''}.
We see no immediate harm in statements `reinforcing' one another like this and adding both forms to the argument graph, but these findings indicate the need for usability testing and trying out alternate visualizations.

\paragraph{Differing Specificity of Similar Statements}
Very similar statements require entirely different arguments.
Yet, by adding arguments, some users were implicitly changing the specificity of the statement they were contributing to. For example, statements regarding second-hand smoke were being raised in a discussion on whether \emph{``Governments should ban smoking''}. This is more relevant in a discussion on whether \emph{``Governments should ban smoking \emph{in public spaces}''}. Such related yet disparate statements need better support to keep discussions more focused.

\paragraph{Comparative Statements}
Many deliberation dialogues---as opposed to information seeking and inquiry dialogs we design for---are comparative in nature, for which the platform is not perfectly suited. For example, the main statement analyzed in the software development blog post was: \emph{``Underscore formatting is more suitable than camel case formatting for code''}.
For the vast majority of underlying supporting/opposing statements, this requires repeating this comparison to ensure statements are `free of context'. For example: \emph{``Underscore formatting of code is faster to read than camel case formatting''}.
On the other hand, these types of discussions are potentially flawed as they often reflect `false dichotomies'; they might rule out other or intermediate options. Emphasizing pairwise comparisons when an argument relies on them has the potential benefit of highlighting the limited scope of statements. 

\paragraph{Dependence of Statements on Context}
More generally, writing statements that are `free of context' is unintuitive and hard. Many of the first statements added by new users were not living up to this guideline.
Suchman argues that \emph{``the communicative significance of a linguistic expression is always dependent on the circumstances of its use''}~\cite{suchman1987plans}, making it impossible to remove all dependencies on external context.
However, in practice it is possible to elaborate on meaning to such a degree that the likelihood of misunderstandings arising due to differences in interpretation becomes minimal.
The problem is this requires excessive reiteration of context in each supporting and opposing statement, which becomes longer and longer as statements relate to more specific situations.
To resolve this, we envision `context tags' which can be added to statements to introduce implicit context with an associated definition.
When adding a related statement they are inherited by default, but can be removed by users in case they feel a statement can be made more generic.

\section{Conclusion}
Overall, we conclude that the structure imposed on argumentation by \systemName{} is a great first step towards accomplishing our goals to make arguments transparent, inspire critical thinking, help find common ground, and be conducive to discussion---at the cost of making it harder to contribute to arguments.
We identified challenges for future research and reported on insights which may be of interest to other designers of argument technologies.
However, the general observation that structuring one's thoughts in `network structures' is challenging is nothing new~\cite{shum2003roots}: 
\begin{quote}
... argument mapping initially feels like learning a new foreign language, and the temptation is to lapse back into more familiar languages (conversational patterns and modes of writing). The tools can be made user friendly, and the notations lightweight and informal, but the human element of the system must co-evolve as well.
\end{quote}
While users should not have to know argumentation theory (none of our respondents did) in order to participate in argumentation, it would be overly optimistic to hope to eliminate the `hard work' that goes into formulating an effective argument.
The very act of identifying statements and relationships between them is what defines 
\emph{critical thinking}.
Therefore, we can only hope to inspire users to engage in and learn this alternate form of literacy by providing better support and making the work more rewarding and fun.

However, this does not mean we intend to replace traditional writing or believe that all arguments should be constructed as argument maps.
\systemName{} is designed to \emph{augment} linear text. Our long-term goal is to be able to highlight statements anywhere on the web (relying on an `open annotation' infrastructure similar to hypothes.is and rbutr) and linking them to structured argumentation.
We feel this could effectively replace traditional comment sections on news articles, blog posts, or even scientific papers, and provide better support for people to critically evaluate, contest, and reason about information online.
At a glance, the popularity of statements and how well-supported or controversial they are could be assessed, and more advanced statistics might be able to indicate how objective or biased an article is.
Looking even further ahead, making arguments more focused and open to falsification has the potential of scaling up governance and by extension the democratic process~\cite{rahwan2017towards}.







\nocite{*} 
\bibliographystyle{ios1}           
\bibliography{bibliography}        

%

\end{document}